\documentstyle[12pt,epsfig]{article}
\def\e{\epsilon}

\def\zcut{z_{\rm cut}}
\def\ycut{y_{\rm cut}}

\def\smin{s_{\rm min}}
\def\as{\left(\frac{\alpha_s}{2\pi}\right)}
\def\aqed{\left(\frac{\alpha e_q^2}{2\pi}\right)}
\def\Dq{D_{q\to \gamma}}
\def\Pqpzero{P^{(0)}_{q\to \gamma}}
\def\Pqpone{P^{(1)}_{q\to \gamma}}
\def\Pqqzero{P^{(0)}_{q\to q}}
\begin{document}
\pagestyle{plain}

\begin{titlepage}
\vspace*{-1cm}
\begin{flushright}
TTP98-38 \\
October 1998 \\
\end{flushright}                                
\vskip 1.cm
\renewcommand{\thefootnote}{\fnsymbol{footnote}}
\begin{center}                                                             
{\Large\bf
Photon Fragmentation at LEP
\footnote{Talk presented at the Workshop on photon interactions and the
  photon structure, Lund, Sweden, September 10-13,1998.}} 
\vskip 1.3cm

{\large A.~Gehrmann--De Ridder}
\vskip .2cm
{\it Institut f\"{u}r Theoretische  Teilchenphysik, Universit\"{a}t
Karlsruhe, \\ D-76128 Karlsruhe, Germany}

\vskip 2.3cm   
\end{center}      
\begin{abstract}
The production of final state photons in hadronic $Z$-boson decays 
can be used to study the quark-to-photon fragmentation function
$D_{q\to \gamma}(z,\mu_{F})$. Currently, two different observables 
are used at LEP to probe this function: the `photon' +~1 jet rate
and the inclusive photon energy distribution. We outline the results 
of a calculation of the `photon' +~1 jet rate 
at fixed ${\cal O}(\alpha \alpha_{s})$, which yield
a next-to-leading order determination of the quark-to-photon 
fragmentation function $D_{q\to \gamma}(z,\mu_{F})$. The resulting 
predictions for the isolated photon rate and the inclusive photon
spectrum at the same, fixed order, are found to be in good agreement 
with experimental data. Furthermore, we outline the main features of
conventional approaches using parameterizations of the resummed
solutions of the evolution equation and  
point out deficiencies of these currently available 
parameterizations in the large $z$-region. We 
finally demonstrate that the ALEPH data on the 
`photon' +~1 jet rate are able to
discriminate between different parameterizations 
of the quark-to-photon fragmentation function, which are equally
allowed by the OPAL photon energy distribution data.
\end{abstract}                                                                
\vfill
\end{titlepage}
\newpage
\renewcommand{\thefootnote}{\arabic{footnote}} 
\setcounter{footnote}{0}
\section{Introduction}

The production of final state photons at large transverse momenta is 
one of the key observables studied in hadronic collisions. 
Data on high-$p_T$ photon production  
yield valuable information on the gluon distribution in the proton,
while the presence of photons in the final state 
represents an important background source in many searches 
for new physics. A good understanding
of direct photon production within the context of 
the Standard model is therefore essential.

Photons produced in hadronic collisions arise essentially 
from two different sources: the {\em direct} production of a 
photon off a primary parton or  
through the {\em fragmentation} of a hadronic jet into a single photon 
carrying a large fraction of the jet energy.
The former gives rise to perturbatively calculable short-distance 
contributions whereas the latter is primarily a long 
distance process which cannot be calculated within perturbative QCD. 
It is described by the process-independent parton-to-photon 
fragmentation function~\cite{phofrag} 
which must be determined from experimental data. 
Its evolution with the factorization scale $\mu_{F}$ can however 
be determined by perturbative methods.

Directly emitted photons are usually well separated 
from all hadron jets produced in a particular event, while photons  
originating from fragmentation processes are primarily to be found 
inside hadronic jets. Consequently, by imposing some isolation 
criterion on the photon, one is in principle able to suppress (but 
not to eliminate) the fragmentation contribution to final state 
photon cross sections, and thus to define {\it isolated} photons. 
However, recent analyses of the production of isolated photons 
in electron-positron and proton-antiproton collisions 
have shown that 
the application of a geometrical isolation cone surrounding the 
photon does not lead to a reasonable agreement between theoretical 
prediction and experimental data. 

An alternative approach to study final state photons produced in 
a hadronic environment is obtained by applying the so-called democratic 
clustering procedure \cite{andrew}. In this approach, the photon is 
treated like any other hadron and is clustered simultaneously 
with the other hadrons into jets. Consequently, one of the 
jets in the final state contains a photon and is labelled `photon jet'
if the fraction of electromagnetic energy within the jet is 
sufficiently large,
\begin{equation}
z=\frac{E_{EM}}{E_{EM} +E_{HAD}}>\zcut,
\label{eq:zdef}
\end{equation}
with $\zcut$ determined by the experimental conditions.
This photon is called  
{\it isolated} if it carries more than a certain fraction, typically 95\%,
of the jet energy and said to be non-isolated otherwise.
Note that this separation is made by studying the experimental $z$ 
distribution and is usually such that hadronisation effects,
which tend to reduce $z$, are minimized.

This democratic procedure has been applied by the ALEPH
collaboration at CERN 
in an analysis of two jet events in electron-positron annihilation 
in which one of the jets contains a highly energetic photon
\cite{aleph}. 
In this analysis, 
ALEPH made a leading order determination of
the quark-to-photon fragmentation function
by comparing the 
photon +~1 jet rate calculated up to ${\cal O}(\alpha)$ 
\cite{andrew} with the data. The theoretical basis on which 
the measurement of the `photon' +~1 jet rate 
relies, is an explicit counting of powers of the strong coupling $\alpha_{s}$ 
in both the direct and the fragmentation contributions, no resummation of $\ln \mu_{F}^2$ is performed.
We shall refer to this theoretical framework as the fixed order
approach. In Section \ref{sec:gammanlo}, we will describe the main features
of the leading and next-to-leading order calculation of the 
photon +~1 jet rate following
this fixed order approach and  see 
how the obtained predictions compare with the available data.

More recently, the OPAL collaboration has measured the inclusive 
photon distribution for final state photons 
with energies as small as 10 GeV \cite{OPAL}.
This corresponds to the photon carrying a fraction of the
beam momentum, $x_{\gamma}$, to be as low as 0.2. 
They have compared their results with the two model-dependent predictions 
of GRV \cite{grv} and BFG \cite{bfg} and found a reasonable agreement 
in both cases when
choosing the factorization scale $\mu_F=M_Z$.
In Section \ref{sec:gammanlo} we shall compare the predictions 
obtained for the inclusive rate within our fixed order approach 
with the OPAL data too.
  
The model predictions \cite{grv,bfg}
mentioned above are based on a resummation of 
the logarithms of the factorization scale $\mu_{F}$ 
and naturally associate an inverse power of $\alpha_{s}$ 
with all fragmentation contributions. We shall
refer to this  resummation procedure 
as the conventional approach. In Section \ref{sec:gammabll} we shall present 
the main features of this approach, outline the results 
obtained for the photon +~1 jet rate in this approach while using either the
GRV or BFG parameterizations for the photon fragmentation function
and show how these compare with the ALEPH data.
Finally Section \ref{sec:Conclusions} contains our conclusions.

\section{The photon +~1 jet rate in the fixed order framework }
\label{sec:gammanlo}
\subsection{The photon +~1 jet rate at ${\cal O}(\alpha)$}

In the fixed order framework, the 
cross section for the production of isolated photons 
receives sizeable contributions from both direct photon and
fragmentation processes. More precisely,
the distribution of electromagnetic energy within 
the photon jet of photon +~1 jet events, for a single quark of 
charge $e_q$, at ${\cal O}(\alpha)$  in the $\overline{{\rm MS}}$-scheme,
can be written as \cite{andrew},
\begin{eqnarray}
\frac{1}{\sigma_0} \frac{{\rm d}\sigma^{(LO)}}{dz}
&=& 2 D_{q\to \gamma}(z,\mu_F)
 + \frac{\alpha e_q^2}{\pi} 
P_{q \gamma}^{(0)}(z) \log \left(\frac{s_{\rm cut}}{\mu_F^2}\right) \,+\,
R_{\Delta}\delta(1-z) + \ldots, \nonumber \\
&=& 2 D_{q\to \gamma}(z,\mu_F) +C_{\gamma}^{(0)}(z,\mu_F).
\label{eq:sig0}
\end{eqnarray}
The $\ldots$ represent terms which are 
well behaved as $z \to 1$.
$C^{(0)}_{\gamma}$ is the coefficient function corresponding to 
the lowest order process $e^+e^- \to q \bar{q} \gamma$.
It is defined after the leading quark-photon singularity 
has been subtracted and factorized in the bare quark-to-photon 
fragmentation function in the $\overline{{\rm MS}}$ scheme.
The non-perturbative fragmentation function 
is an exact solution at ${\cal O}(\alpha)$ of 
the evolution equation in the factorization scale $\mu_{F}$,
\begin{equation}
D_{q\to \gamma}(z,\mu_{F}) = 
\frac{\alpha e_q^2}{2 \pi} P_{q \gamma}^{(0)}(z)
\log\left(\frac{\mu_{F}^2}{\mu_{0}^2}\right) + D_{q\to \gamma}(z,\mu_{0}).
\end{equation}
In this equation, all unknown long-distance effects are related 
to the behaviour 
of $D_{q\to \gamma}(z,\mu_{0})$, the initial value of 
this fragmentation function which has been fitted to the data 
at some initial scale $\mu_{0}$ in~\cite{aleph}.
As $D_{q\to \gamma}(z,\mu_{F})$ is exact, 
this solution does not take the commonly implemented 
\cite{grv,bfg} resummations of $\log(\mu_{F}^2)$ into account and when 
used to evaluate the photon +~1 jet rate at ${\cal O}(\alpha)$ yields a 
factorization scale independent prediction for the cross section. 

In the Durham jet algorithm, $s_{\rm cut} 
\sim sz(1-z)^2/(1+z) \sim p_T^2$ \cite{durham} where $p_T$ is the 
transverse momentum of the photon with respect to the cluster.
For $z<1$, we find that $\mu_F^2 \sim s_{\rm cut}$   
and $\mu_F^2 \gg \mu_{0}^2$ . The `direct' contribution 
in eq.~(\ref{eq:sig0}) is therefore suppressed relative 
to the fragmentation contribution. 
The conventional assignment of a power 
of $1/\alpha_s$ to the fragmentation function can in this case be  
motivated, this contribution is indeed more significant. 
However, as $z\to 1$, 
we see that the transverse size of the photon jet cluster 
decreases such that $s_{\rm cut}\to 0$.
The hierarchy $s_{\rm cut} \sim \mu_F^2$ and $\mu_F^2 \gg \mu_0^2$ is
no longer preserved and both contributions in
eq.~(\ref{eq:sig0}) are important.
Large logarithms of $(1-z)$ become the most dominant contributions.
Being primarily interested in the high $z$ region, in \cite{andrew}
it was chosen not to impose the conventional prejudice 
 and resum the logarithms of $\mu_F$ {\em a priori} but to work 
within a fixed order framework, to isolate the relevant large logarithms.  

We have performed the calculation of the
${\cal O}(\alpha_s)$ corrections to the 
`photon' +~1 jet rate using the same democratic procedure 
to define the photon as in~\cite{andrew,aleph}.
The details of this fixed order calculation 
have been presented in ~\cite{big}.  In the following, 
we shall limit ourselves to outline the main characteristics 
of this calculation, 
to summarize the results and 
to show how these compare with the available experimental data from ALEPH.

%%%%%%%%%%%%%%%%%%%%%%%%%%%%%%%%%%%%%%%%%%%%%%%%%%%%%%%%%%%%%%%%%%%%%
\subsection{The `photon' +~1 jet rate at  ${\cal O}(\alpha \alpha_s)$}
%%%%%%%%%%%%%%%%%%%%%%%%%%%%%%%%%%%%%%%%%%%%%%%%%%%%%%%%%%%%%%%%%%
\begin{figure}[t]
\begin{center}
~ \epsfig{file=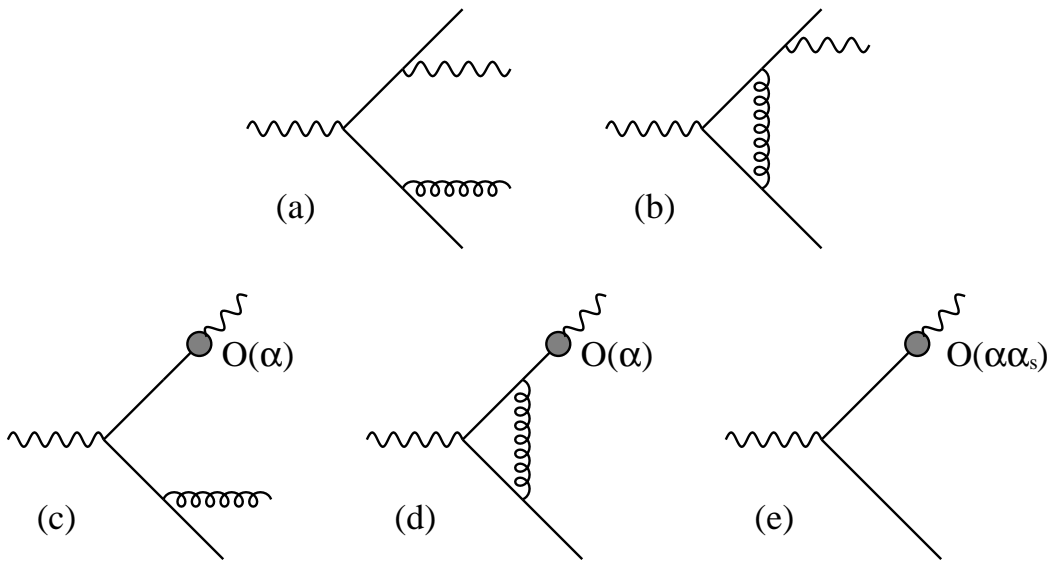,width=14cm}
\caption{Parton level subprocesses contributing to the photon +~1 jet 
rate at ${\cal O}(\alpha\alpha_s)$.} 
\label{fig:class}
\end{center}
\end{figure}
The `photon' +~1 jet rate in $e^+e^-$ annihilation at ${\cal O} (\alpha 
\alpha_s)$ receives contributions from five parton-level subprocesses
displayed in Fig.~\ref{fig:class}.
Although the `photon' +~1 jet cross section 
is finite at ${\cal O}(\alpha \alpha_{s})$, 
all these contributions  
contain divergences (when the photon 
and/or the gluon are collinear with one of the quarks, 
when the gluon is soft or since the bare quark-to-photon fragmentation 
function contains infinite counter terms).
All these divergences have to be isolated and cancelled analytically
before the `photon' +~1 jet cross section can be evaluated numerically.

\begin{figure}[t]
\vspace{10cm}\begin{center}
~ \includegraphics{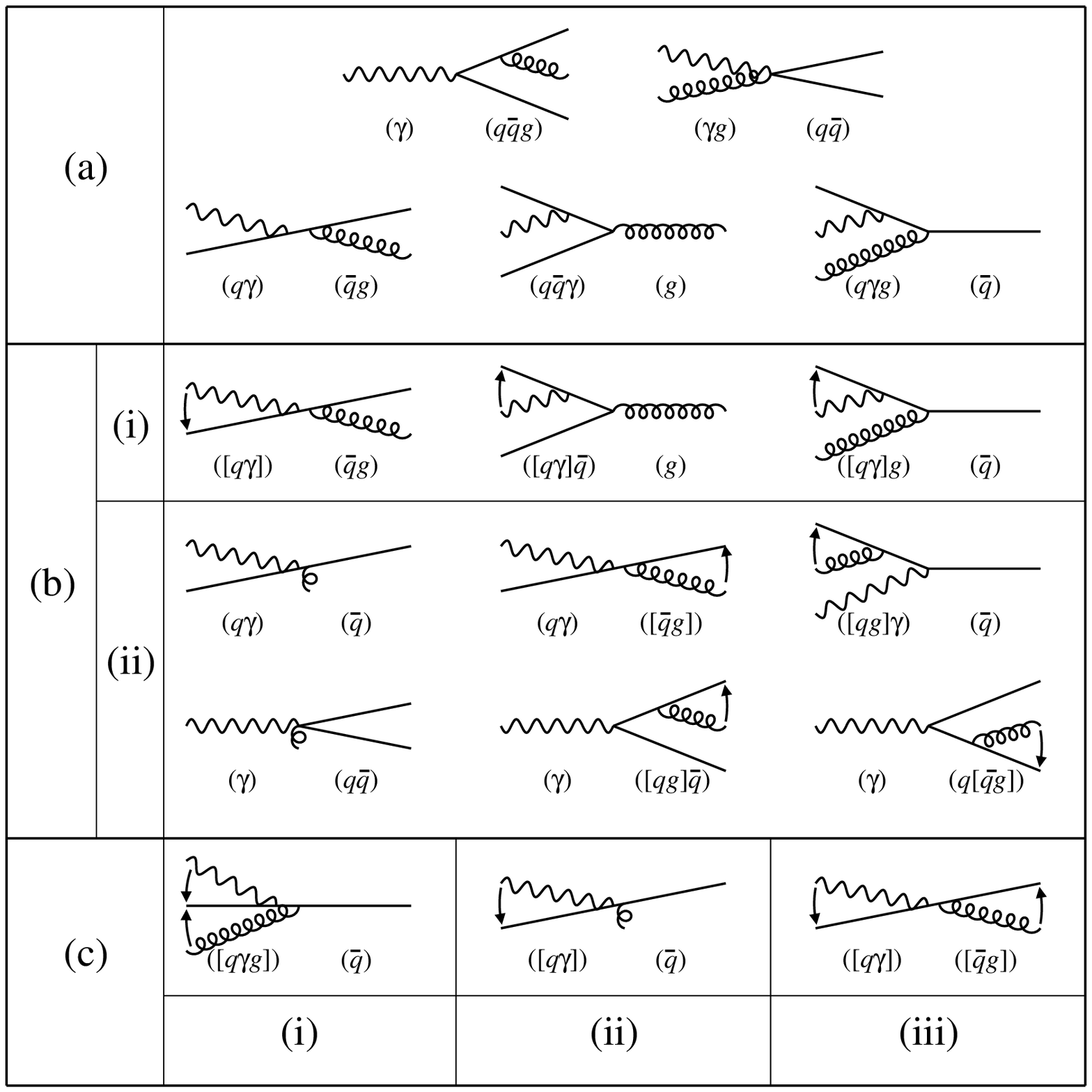}
\caption{Different final state `photon' + 1~jet topologies arising  
from the tree level  $\gamma^* \to q
\bar{q}\gamma g$ process.  The `photon' jet is moving to the left while the recoiling hadronic jet moves to the right.
Square brackets denote theoretically
unresolved particles, round brackets experimental clusters.}
\label{fig:clus4}
\end{center}
\end{figure}

The various configurations where the tree level process
 $\gamma^* \to q \bar{q}g\gamma$ contributes
to the photon +~1 jet rate are illustrated in Fig.~\ref{fig:clus4}.
The associated real contributions can be separated into three
 categories:  either
theoretically resolved, single unresolved or double unresolved.
Within each singular region which is defined 
using a theoretical criterion $\smin$, 
the matrix elements are approximated and 
the unresolved variables analytically integrated.
The evaluation of the singular contributions 
associated with the process $\gamma ^* \to q \bar{q}g \gamma$ is of 
particular interest as it contains various ingredients 
which could directly be applied to the calculation of jet observables 
at next-to-next-to-leading order.
Indeed, besides the contributions arising when 
one final state gluon is collinear or soft (single unresolved
 contributions, see fig.\ref{fig:clus4}(b)), 
there are also contributions where {\em two} of the 
final state partons are theoretically unresolved, see fig.\ref{fig:clus4}(c) .
The three different double unresolved contributions which 
occur in this calculation are: 
the {\it triple collinear} contributions, 
arising when the photon and the gluon are simultaneously collinear 
to one of the quarks, the {\it soft/collinear} contributions 
arising when the photon is collinear to one of the quarks 
while the gluon is soft and the {\it double single collinear} contributions, 
resulting when the photon is collinear to one of the quarks while the gluon 
is collinear to the other.
A detailed derivation of each of these singular real contributions 
and of the singular contributions arising in the processes depicted 
in Fig.~\ref{fig:class}(b)-(d) has been presented in~\cite{big}. 

Combining all unresolved contributions present in the 
processes shown in Fig.~\ref{fig:class}(a)-(d)  
yields a result 
that still contains single and double poles in $\e$.
These pole terms are however proportional 
to the universal next-to-leading order splitting function 
$P_{q\gamma}^{(1)}$ ~\cite{curci} and a convolution of two 
lowest order splitting functions, 
$(P_{qq}^{(0)}\otimes P_{q \gamma}^{(0)})$. 
Hence, they can be factorized into the next-to-leading order 
counterterm of the bare quark-to photon fragmentation function 
\cite{fac} present in the contribution depicted in Fig.~\ref{fig:class}(e),
yielding a finite and factorization scale ($\mu_{F}$) dependent 
result \cite{big}.

We have then chosen to evaluate the remaining 
finite contributions numerically using the {\it hybrid subtraction} method, 
a generalization of the {\it phase space slicing} procedure \cite{gg,kramer}.  
The latter procedure turns out to be inappropriate   
when more than one particle is potentially unresolved.
Indeed, in our calculation we found areas in the four parton phase space 
which belong simultaneously to two different single collinear regions.
Those areas cannot be treated correctly within the phase space 
slicing procedure.
Within the {\it hybrid subtraction} method developed in \cite{eec}, 
a parton resolution criterion $\smin$ 
is used to separate the phase space into different resolved and 
unresolved regions.
Phase space slicing and hybrid subtraction methods vary only 
in the numerical treatment of the unresolved regions.
While the matrix elements are set to zero in the former method, one
considers the difference between the full matrix 
element and its approximation in all unresolved regions in the latter.
The non-singular contributions 
are calculated using Monte Carlo methods like in the phase space 
slicing scheme.
 
The numerical program finally evaluating the `photon' +~1 jet rate 
at ${\cal O}(\alpha\alpha_s)$ contains four separate contributions.
\begin{figure}[t]
\begin{center}
~ \epsfig{file=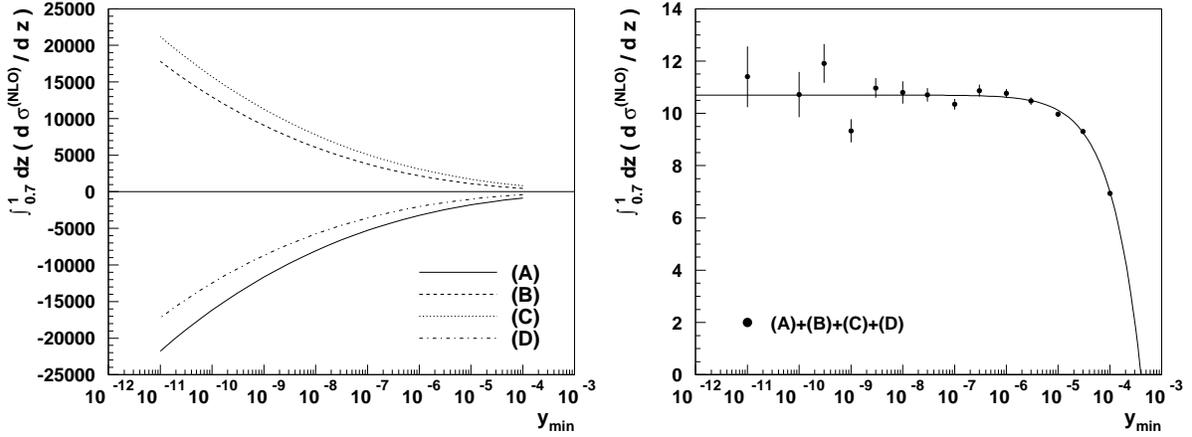,angle=-90,width=16cm}
\caption{${\cal O}(\alpha\alpha_s)$ individual contributions (left) 
and sum of all ${\cal O}(\alpha\alpha_s)$ contributions (right) to the 
photon +~1 jet rate for a single quark of charge $e_q$ such that
$\alpha e_q^2 = 2\pi$, $\alpha_s (N^2-1)/2N = 2\pi$ using
the Durham  jet algorithm with $y_{{\rm cut}}=0.1$,  
and integrated for $z>0.7$.}
\label{fig:ymin2}
\end{center}
\end{figure}
Each of them depends logarithmically (in fact 
as $\log^3 (y_{{\rm min}})$) on 
the theoretical resolution parameter $y_{{\rm min}}=s_{{\rm min}}/Q^2$. 
The physical `photon' + 1~jet cross section,
which is the sum of all four contributions,
{\it must} of course be independent of the choice of $y_{{\rm min}}$,
the latter being just an artefact of the theoretical calculation.
In Fig.~\ref{fig:ymin2}, we see that the cross section approaches (within 
numerical errors) a constant value provided that $y_{{\rm min}}$ 
is chosen small enough, indicating 
that a complete cancellation of all powers of $\log (y_{{\rm min}})$ 
takes place. This 
provides a strong check on the correctness of our results 
and on the consistency of our approach.

Finally, after factorization of the quark-photon singularities, the 
${\cal O}(\alpha \alpha_s)$ cross section takes the following form, 
\begin{equation}
\frac{1}{\sigma_{0}}\frac {{\rm d} \sigma^{(NLO)}}{{\rm d}z}
=
\frac{1}{\sigma_0}\frac{{\rm d}\sigma^{(LO)}}{{\rm d}z}
+ \as \aqed
C^{(1)}_{\gamma}(z,\mu_{F})+ 
C_{q}^{(0)}
\otimes D_{q\to \gamma}(z,\mu_{F}).
\label{eq:signlo}
\end{equation}
The lowest order cross section has been given in eq.(\ref{eq:sig0}) while
the hard scattering coefficient functions $C_{i}^{(n)}$ appearing
explicitly in this equation are defined as follows.
The (finite) next-to-leading order coefficient function
$C^{(1)}_{\gamma}$ is obtained numerically 
after the next-to-leading quark-photon singularity has been subtracted.    
More precisely, $C^{(1)}_{\gamma}$ is obtained after summing 
all contributions which are independent 
of $D_{q\to \gamma}(z,\mu_{F})$ 
arising from the Feynman diagrams depicted 
in Fig.~\ref{fig:class} together. 
A detailed description of the evaluation of $C^{(1)}_{\gamma}$ has been given  
in \cite{big}.
The coefficient function $C_{q}^{(0)}$ is the finite part 
associated with the sum of real and virtual gluon contributions 
to the process $e^+e^- \to q \bar q$. It is straightforward to evaluate, 
and can be found for example in \cite{kt}.

\subsection{Comparison with Experimental Data  }

A comparison between the measured `photon' +~1 jet rate  \cite{aleph}
and our calculation yielded a first determination of 
the quark-to-photon fragmentation function accurate up to 
${\cal O}(\alpha \alpha_s)$ \cite{letter}.
This function, which parameterizes the 
perturbatively incalculable long-distance effects, 
has to satisfy a perturbative evolution equation in the factorization 
scale $\mu_F$.
At next-to-leading order (${\cal O}(\alpha \alpha_{s})$) 
this equation reads,
\begin{equation}
\frac{\partial \Dq(z,\mu_{F})}
{\partial \log(\mu_{F}^2)}=
\aqed \Pqpzero(z) \,+\,\aqed \as \Pqpone(z)
+\as \Pqqzero\otimes \Dq(z,\mu_{F}).
\label{eq:evolnlo}
\end{equation}
$\Pqqzero$ and $\Pqpone$ are respectively the lowest order quark-to-quark 
and the next-to-leading order quark-to-photon universal splitting 
functions \cite{curci,rijken,AP}.
The next-to-leading order fragmentation
function can be expressed as an {\it exact} solution of 
this evolution equation 
up to ${\cal O}(\alpha \alpha_s)$ \cite{big},
\begin{eqnarray}
\Dq(z,\mu_{F})&=&
\frac{\alpha e_{q}^2}{2\pi}P^{(0)}_{q \gamma}(z)
\log\left(\frac{\mu^2_{F}}{\mu_{0}^2}\right) 
+\frac{\alpha e_{q}^2}{2\pi} \frac{\alpha_{s}}{2 \pi}
\left(\frac{N^2 -1}{2N}\right)P_{q \gamma}^{(1)}(z)
\log \left(\frac{\mu^2_{F}}{\mu_{0}^2}\right)
\nonumber\\
& & +
\frac{\alpha_{s}}{2 \pi}
\left(\frac{N^2 -1}{2N}\right)
\log \left(\frac{\mu^2_{F}}{\mu_{0}^2}\right) P_{qq}^{(0)}(z)\otimes 
\frac{\alpha e_{q}^2}{2 \pi}\frac{1}{2}P_{q \gamma}^{(0)}(z)
\log \left(\frac{\mu^2_{F}}{\mu_{0}^2}\right)
\nonumber\\
& &
+\frac{\alpha_{s}}{2 \pi}
\left(\frac{N^2 -1}{2N}\right)
\log \left(\frac{\mu^2_{F}}{\mu_{0}^2}\right) P_{qq}^{(0)}(z)\otimes 
D(z,\mu_{0}) \,+D(z,\mu_{0}).
\label{eq:Dnlo}
\end{eqnarray}
The initial function $\Dq^{(NLO)}(z,\mu_{0})$ has been fitted to the ALEPH  ~1 jet
data \cite{letter} for  
$\frac{1}{\sigma_{0}}\frac{d\sigma}{dz}$,  
for the jet resolution parameter $y_{{\rm cut}}=0.06$ 
and $\alpha_s(M_z^2) = 0.124$
to yield \footnote{Note that the logarithmic term proportional to 
$P^{(0)}_{q \gamma}(z)$
is introduced to ensure that the predicted $z$ distribution is 
well behaved as $z \to 1$ \cite{andrew}.}, 
\begin{equation}
D^{(NLO)}(z,\mu_{0})=\frac{\alpha e_{q}^2}{2 \pi} 
\left(-P^{(0)}_{q \gamma}(z)\log(1-z)^2 \;+\,20.8\,(1-z)-11.07\right),
\label{eq:fitnlo}
\end{equation}
where $\mu_{0}=0.64$~GeV.
The next-to-leading order ($\overline{{\rm MS}}$)
quark-to-photon fragmentation function (for a quark of unit charge) 
at a factorization scale $\mu_F=M_Z$ were shown 
in \cite{letter} and 
compared with the lowest order fragmentation function obtained 
in~\cite{aleph}. A large difference between the leading and next-to-leading 
order quark-to-photon fragmentation functions was observed only for  
$z$ close to 1, indicating the presence of large $\log (1-z)$ terms. 

Moreover, a comparison between 
the ALEPH data and the results of the ${\cal O}(\alpha \alpha_s)$ 
calculation using the fitted next-to-leading order 
fragmentation function for different values of $\ycut$ 
can be found in \cite{big,letter}.
The next-to-leading order corrections were found to be moderate 
for all values of $\ycut$ demonstrating the 
perturbative stability of our fixed order approach.
\begin{figure}[t]
\begin{center}
~ \epsfig{file=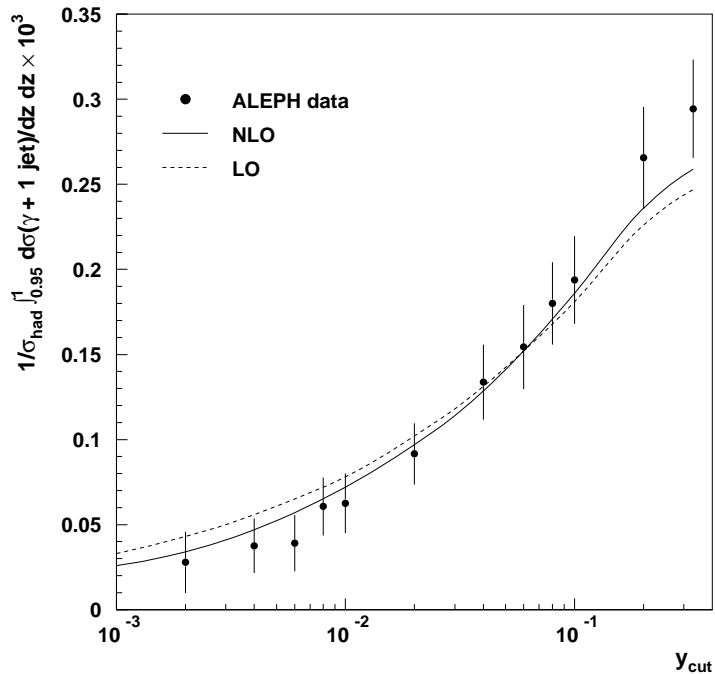,width=9cm,angle=-90}
\caption{The integrated photon +~1 jet rate above $z=0.95$ as function of
$y_{{\rm cut}}$, compared with the ${\cal O}(\alpha)$ 
and ${\cal O}(\alpha \alpha_s)$
order calculations including the appropriate 
quark-to-photon fragmentation functions.}
\label{fig:ycut}
\end{center}
\end{figure}
To test the generality of our results, we have considered 
two further applications: 
the `isolated' photon rate and the inclusive photon distribution 
which we shall now briefly present.
  
Using the results of the calculation of the photon +~1 jet rate at 
${\cal O}(\alpha \alpha_s)$ and  
the fitted quark-to-photon fragmentation function, we have 
determined the {\it isolated} 
rate defined as the photon +~1 jet rate for $z>0.95$ 
in the democratic approach. 
The result of this calculation compared with 
data from ALEPH~\cite{aleph} and the leading order calculation~\cite{andrew}
is shown in Fig.~\ref{fig:ycut}. It can clearly be seen that inclusion of 
the next-to-leading order corrections improves the agreement between 
data and theory over the whole range of $y_{{\rm cut}}$. 
It is also apparent that the next-to-leading order corrections 
to the isolated photon +~1 jet rate obtained in this 
democratic clustering approach are of reasonable size 
indicating a good perturbative stability of this {\it isolated} 
photon definition. 
\begin{figure}[t]
\begin{center}
~ \epsfig{file=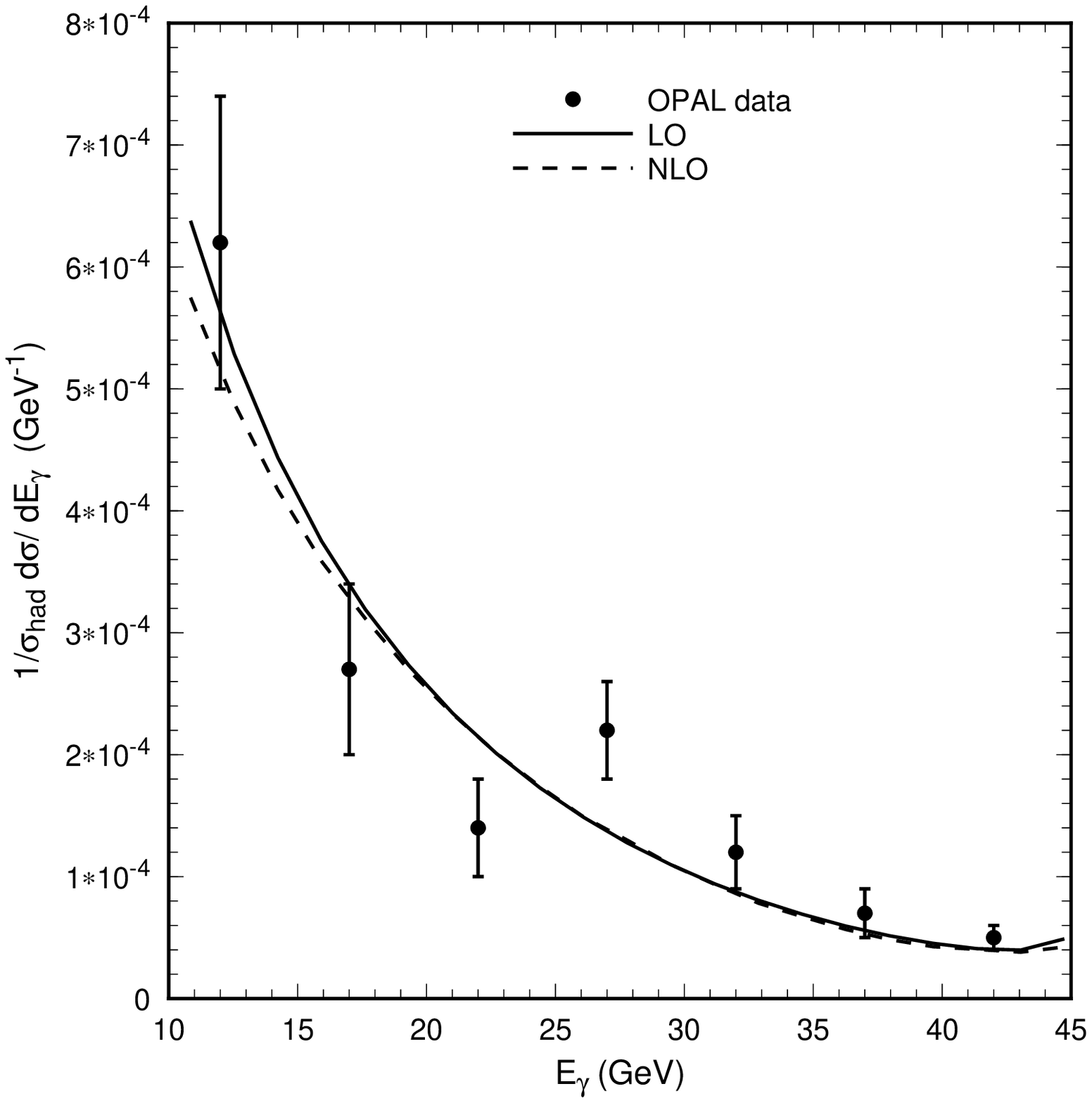,width=9cm}
\caption{The inclusive photon energy distribution 
normalized to the hadronic cross section as measured by the OPAL Collaboration
compared with the ${\cal O}(\alpha)$ and ${\cal O}(\alpha \alpha_s)$
order calculations including the appropriate 
quark-to-photon fragmentation functions 
determined using the ALEPH photon +~1 jet data.}
\label{fig:inclusive}
\end{center}
\end{figure}

The OPAL collaboration has recently measured the inclusive photon distribution
for final state photons with energies between 10 and 42~GeV \cite{OPAL}.
 Fig.~\ref{fig:inclusive} shows our (scale independent) predictions 
for the inclusive photon energy distribution at both leading and 
next-to-leading order.
We see good agreement with the data, even though the phase space 
relevant for the OPAL data, which corresponds to $x_{\gamma}$ 
values as small as 0.2, 
far exceeds that used to determine the fragmentation functions from the ALEPH
photon +~1 jet data.

%%%%%%%%%%%%%%%%%%%%%%%%%%%%%%%%%%%%%%%%%%%%%%%%%%%%%%%%%%%%%%%%%%%%%%%%%%
\section{The photon +~1 jet rate in the conventional approach}
\label{sec:gammabll}
%%%%%%%%%%%%%%%%%%%%%%%%%%%%%%%%%%%%%%%%%%%%%%%%%%%%%%%%%%%%%%%%%%%%%%%%%%%%

In the conventional approach, the parton-to-photon fragmentation function
$D_{i\to \gamma}$ satisfies an all order inhomogeneous
evolution equation \cite{AP}.
Usually, these equations can be diagonalized in 
terms of the singlet and non-singlet
quark fragmentation functions as well as the gluon fragmentation function.
However when analyzing the global features of the solutions 
of these evolution equations, as was discussed in \cite{papernew},
several simplifications can be consistently made. 
For example, the gluon-to-photon fragmentation function is by orders 
of magnitude smaller than the quark-to-photon fragmentation functions,
as was shown in \cite{papernew}. Its contribution to the photon 
cross section can safely be ignored. 
Consequently, the flavour singlet and 
non-singlet quark-to-photon fragmentation functions become equal to a unique 
fragmentation function $\Dq$ which satisfies 
an evolution equation having a similar form than  
the next-to-leading order evolution 
valid in the fixed order approach, see eq.~(\ref{eq:evolnlo}).
Unlike in 
eq.~(\ref{eq:evolnlo}) though, the strong coupling $\alpha_{s}$
is now a function of the factorization scale, it runs. 

The full solution $\Dq$ of the inhomogeneous evolution equation is given by  
the sum of two contributions; a pointlike (or perturbative) part $\Dq^{pl}$
which is a solution of the inhomogeneous equation follows 
eq.(\ref{eq:evolnlo}) 
and a hadronic (or non-perturbative) part $\Dq^{had}$ 
which is the solution of the corresponding homogeneous equation.
In the conventional approach, approximate solutions of these evolution 
equations are commonly obtained as follows \cite{grv,bfg}. 
First an analytic solution in moment space is 
obtained in the leading logarithm (LL) or beyond leading logarithm (BLL) 
approximations. These are then inverted numerically to give 
the fragmentation function in $x$-space.
At LL only terms of the form $(\alpha_s^n \ln^{n+1} \mu_F^2)$ 
are kept while at BLL both leading
$(\alpha_s^n \ln^{n+1} \mu_F^2)$ and subleading  $(\alpha_s^n \ln^n \mu_F^2)$
logarithms of the mass factorization scale $\mu_F$ are resummed 
to all orders in the strong coupling $\alpha_{s}$.

It is worth noting that  both LL and BLL solutions have an asymptotic 
behaviour given by,
\begin{equation}
\Dq^{asympt}(z,\mu_F)=\aqed \frac{2 \pi}{\alpha_{s}(\mu_{F}^2)}\,a(z),
\label{eq:asympt}
\end{equation}
where $a(z)$ contains the splitting function $\Pqpzero$.
This asymptotic form lends support to
the common assumption that the
quark-to-photon fragmentation function $\Dq$ is   
${\cal O}\left( \alpha/\alpha_{s} \right)$. 
This assumption is in contrast with 
that adopted in the fixed order approach (cf. Section~\ref{sec:gammanlo}) 
where the quark-to-photon fragmentation function is  ${\cal O}(\alpha)$.
It leads  to significant differences in the respective expressions 
of the one-photon production cross sections. 
Indeed, the  LL and BLL expression of the cross section in the 
$\overline{{\rm MS}}$ scheme arising when one uses the corresponding 
resummed LL or BLL fragmentation functions in this approach 
reads
\begin{eqnarray}
\frac{1}{\sigma_{0}}\frac {{\rm d} \sigma^{LL}}{{\rm d}z} &=&
\;D_{q\to \gamma}(z,\mu_{F}), 
\nonumber\\
\frac{1}{\sigma_{0}}\frac {{\rm d} \sigma^{BLL}}{{\rm d}z}
&=&
D_{q\to \gamma}(z,\mu_{F})\,
+\as C_{q}^{(0)}
\otimes D_{q\to \gamma}(z,\mu_{F}) 
+ \aqed  C^{(0)}_{\gamma}(z,\mu_{F}).
\label{eq:sigbll}
\end{eqnarray}
As can be seen from these equations, no direct term contributes to the
cross section at the LL level, while only the ${\cal O}(\alpha)$ direct
term $C^{(0)}_{\gamma}$ contributes at the BLL level to it. 
At the BLL level, contributions arising
from \ref{fig:class}(a)-(b) do not enter in the cross section.
As explained at length in~\cite{letter}, this conventional 
procedure of associating an inverse power of $\alpha_{s}$ 
with the fragmentation function 
is clearly appropriate when the logarithms of the factorization scale 
$\mu_{F}$ are the {\em only} potentially large logarithms but 
is problematic when different classes of large 
logarithms can occur as it is the case in the `photon' +~1 jet cross section.

All unknown long-distance effects are
related to the behaviour of the input fragmentation function 
$D_{q\to \gamma}^{np}(z,\mu_{0})$ implicitly present in eq.(\ref{eq:sigbll}).
In the approaches of GRV or BFG, 
the non perturbative input function 
$D_{q \to \gamma}^{np}(z,\mu_{0})$ is treated with only 
minor differences. Those have been detailed in \cite{papernew}.
We shall here concentrate on the major common points in these approaches.
At LL both GRV and BFG agree  
that $D_{q \to \gamma}^{np}(z,\mu_{0})$
is negligible and can be described by a vector meson dominance model (VMD) 
as explained in \cite{grv} and \cite{bfg} respectively.
However at BLL and in the $\overline{{\rm MS}}$ scheme, 
the input fragmentation function cannot be negligible 
due to the presence of the direct term $C_{\gamma}^{(0)}$  and 
cannot be described by a VMD  input alone. 
Indeed, $C_{\gamma}^{(0)}(z)$ diverges as $z \to 1$ and would drive 
the cross section to unacceptable negative values if a VMD input 
alone is considered for the input fragmentation function.
Note that the requirement that the
cross section is positive led the authors in \cite{andrew, letter} to consider 
in the fixed order approach a term proportional to $\Pqpzero \ln(1-z)^2$
in the expression of $D_{q \to \gamma}^{np}(z,\mu_{0})$.
In summary, in any resummed or fixed order approach, as soon 
as the direct term $C_{\gamma}^{(0)}$ enters the cross section,
as it does in the $\overline{{\rm MS}}$ factorization scheme, 
the input fragmentation function 
$D_{q \to \gamma}^{np}(z,\mu_{0})$ must compensate the large $z$
behaviour of $C_{\gamma}^{(0)}$. Consequently, 
this input fragmentation function 
$D_{q \to \gamma}^{np}(z,\mu_{0})$ as well as the total solution 
$D_{q \to \gamma}(z,\mu_{F})$ in this $\overline{{\rm MS}}$ scheme 
should clearly exhibit a divergent behaviour as $z \to 1$ .

\begin{figure}[t]
\begin{center}
~ \epsfig{file=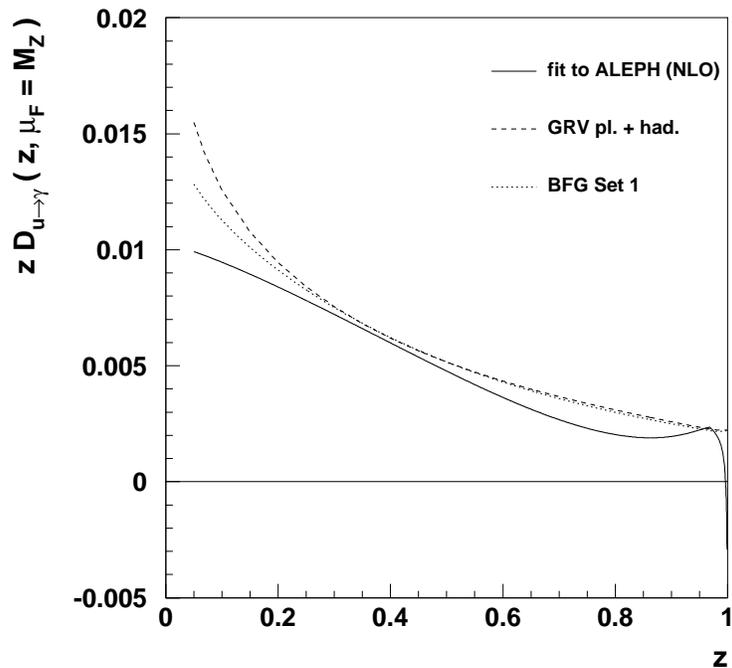,width=9cm,angle=-90}
\caption{The quark-to-photon fragmentation function 
$z D_{u \to \gamma}(z,\mu_F)$  
evaluated at $\mu_F = M_Z$ in the $(\overline{{\rm MS}})$-scheme. 
The NLO fit from the ALEPH `photon' + 1~jet data is shown
as solid line.  The  GRV (BFG) parameterization is shown
dashed (dotted).} 
\label{fig:dfrag}
\end{center}
\end{figure}

In Fig. \ref{fig:dfrag} we compare the analytic expression of the fragmentation
function obtained in the fixed order approach, eq.(\ref{eq:Dnlo}) with the BLL
parameterizations of GRV and BFG for the numerically resummed solutions.
We clearly see, that the fixed order solution 
does diverge as $z \to 1$ while the numerical solutions do not. 
This significant disagreement is mainly due to
deficiencies in the numerical parameterizations.
In fact, it can be traced back to the presence
of logarithms of $(1-z)$ that are explicit in the expanded result. 
These logarithms should also be present in the numerical resummed results.
However, the parameterizations are necessarily
obtained by inverting only a finite number of moments
and it is a well known problem to describe a logarithmic behaviour with a 
polynomial expansion.
This clearly indicates that the presently available  
parameterizations for the resummed 
fragmentation functions are not accurate at large $z$ and particularly 
for $z > 0.95$.

Except in the very high $z$ region however,
we see that, the various fragmentation functions generally agree 
well with each other in shape and magnitude.
Consequently we can expect, that predictions 
for the inclusive photon cross sections
(which run over a wide range of $z$) will be largely in agreement, while
significant differences may be apparent in the `photon' + 1~jet
estimates which focus on the large $z$ region.
Indeed, we mentioned that the OPAL data were in agreement with predictions 
using either (GRV or BFG) parameterizations in the conventional
approach, and we showed in Section \ref{sec:gammanlo} that these data were
also in agreement with the predictions obtained in our fixed order approach.

Let us now concentrate on the `photon' +~1 jet cross section, an
observable which is sensitive on the large $z$ region ($0.7<z<1$) and see how the
predictions obtained in a conventional approach compare with 
the ALEPH data. In the following, we focus on  
one particular value of the jet clustering parameter $\ycut$,
$\ycut=0.1$.
\begin{figure}[t]
\begin{center}
~ \epsfig{file=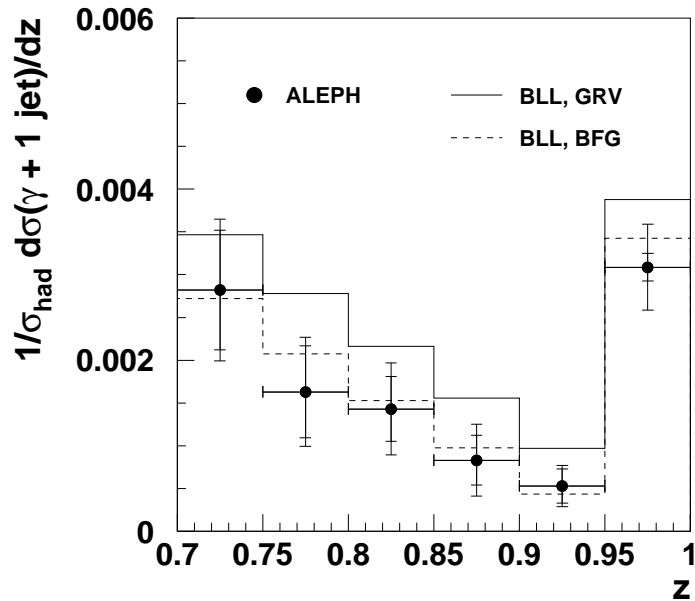, width=8cm,angle=-90} 
\caption{The `photon' + 1~jet rate in the conventional
  approach using the GRV or the BFG inputs.
The experimental data is taken from \cite{aleph}.}
\label{fig:jetfinal}
\end{center}
\end{figure} 
Fig.~\ref{fig:jetfinal} shows the BLL
`photon' + 1~jet rates  obtained 
using either the GRV or BFG parameterizations of the photon
fragmentation function 
for $\mu_{F}=M_{Z}$. 
Ignoring the large $z$ region where we have reason to
doubt the accuracy, 
we see that the BFG predictions lie systematically below that obtained using 
the GRV parameterization and go through the experimental data points.
As discussed in \cite{papernew}, 
this difference is due to both the choice of hadronic scale and 
the non-VMD input.  The BFG input is smaller and the `photon' + 1~jet
data clearly selects this choice.
Notice however, that the BFG parameterization for the fragmentation function 
unlike that of the GRV group was proposed well after 
the ALEPH data were released.

%%%%%%%%%%%%%%%%%%%%%%%%%%%%%%%%%%%%%%%%%%%%%%%%%%%%%%%%%%%%%%%%%%%%%%%%%%%%%

\section{Conclusions}
\label{sec:Conclusions}
%%%%%%%%%%%%%%%%%%%%%%%%%%%%%%%%%%%%%%%%%%%%%%%%%%%%%%%%%%%%%%%%%%%%%%%
In summary, in Section \ref{sec:gammanlo} we have outlined the main features 
of the calculation~\cite{big} 
of the `photon' +~1 jet rate at ${\cal O}(\alpha\alpha_s)$. Although 
only next-to-leading order in perturbation theory, this calculation 
contains several ingredients appropriate to the calculation of 
jet observables at next-to-next-to-leading order. In particular, it requires  
to generalize the phase space slicing method of~\cite{gg,kramer}
to take into account contributions where more than one 
theoretically unresolved particle is present in the final state.
The `photon' +~1 jet rate has then been evaluated 
for a democratic clustering algorithm with a Monte Carlo program using 
the hybrid subtraction method of~\cite{eec}.
The results of our calculation, when compared to the data~\cite{aleph}
on the `photon' +~1 jet rate obtained by ALEPH, 
enabled a first determination
of the process independent quark-to-photon fragmentation function
at ${\cal O}(\alpha\alpha_s)$ in a fixed order approach. 
As a first application, we have used this 
function to calculate the `isolated' photon +~1 jet rate in a democratic 
clustering approach at next-to-leading order. The inclusion of the QCD
corrections improves the agreement between theoretical
prediction and experimental data.
Moreover, it was shown that these corrections are moderate,
demonstrating 
the perturbative stability of this particular isolated photon definition.
We have also shown that the inclusive 
photon energy distribution computed in this fixed order approach and
using the quark-to-photon fragmentation function determined with the
ALEPH data and is in good agreement with the recent OPAL data.

In Section \ref{sec:gammabll} we have outlined the main
characteristics of the conventional approach have described how the
LL and BLL solutions of the evolution equation are obtained.
An important feature characterizing this
approach concerns the power of $\alpha_{s}$ associated to the 
photon fragmentation function.
Although this appears to be ${\cal O}(\alpha)$, 
inspection of the evolution equation suggests
a logarithmic growth with $\mu_F$, and in conventional approaches, 
it is ascribed a nominal power of $\alpha/\alpha_s$.  
Consequences for the LL and BLL cross sections were 
also discussed in this section.

The full $\mu_F$-dependent solution $D_{q \to \gamma}(z,\mu_{F})$ 
of the evolution can
only be obtained provided some non-perturbative input is given.  
In the approaches of GRV or BFG,
this input has two pieces, a small vector meson dominance contribution 
together with a perturbative counterterm. 
In either case,
we have found that the large $z$ behaviour of the 
fragmentation functions is not well reproduced by the 
parameterizations, the main problem being to describe 
a logarithmic behaviour with a polynomial.
Predictions using any of the presently available (GRV or BFG) 
resummed fragmentation functions do therefore 
not yield accurate results in the region $z>0.95$. 
Ignoring this high $z$ region, the BLL predictions for the photon +~1
jet rate obtained using the BFG parameterization are found to be 
in agreement with the ALEPH data, while the predictions using the
fragmentation function of GRV lie too high.

To summarize,
we have seen that the inclusive and `photon' +~1 jet data from LEP can 
be described using either the ${\cal O}(\alpha \alpha_{s})$ 
fragmentation function 
whose non-perturbative input is fitted to the
ALEPH data or using the BLL parameterization of BFG.
In the latter case, 
the agreement needs however to be restricted to $z$-values below 0.95. 

\section*{Acknowledgements}
I wish to thank G.~Jarlskog, L.~J\"onsson and T.~Sj\"ostrand for
organizing an interesting and pleasant workshop.

\end{document}